\documentclass[aps,prl,twocolumn,groupedaddress]{revtex4}
\usepackage{amsmath,amssymb}
\usepackage{graphicx}
\usepackage{bm}
\begin{document}


\title{Coherent manipulation of electron spins up to ambient temperatures in Cr$^{5+}$(S=1/2) doped K$_3$NbO$_8$}

\author{S. Nellutla,$^1$ K.-Y. Choi,$^{1,2}$ M. Pati,$^2$ J. van Tol,$^{1,2}$ I. Chiorescu,$^{1,3}$
and N. S. Dalal$^{1,2}$}

\affiliation{$^1$ National High Magnetic Field Laboratory,
Tallahassee, Florida 32310, USA \\
$^2$ Department of Chemistry and Biochemistry, Florida State
University, Tallahassee, Florida 32306-4390, USA \\
$^3$ Department of Physics, Florida State University, Tallahassee,
Florida 32306-4350, USA}




\date{\today}

\begin{abstract}
We report coherent spin manipulation on Cr$^{5+}$ (\emph{S} = 1/2,
\emph{I} = 0) doped K$_3$NbO$_8$, which constitutes a dilute
two-level model relevant for use as a spin qubit. Rabi oscillations
are observed for the first time in a spin system based on transition
metal oxides up to room temperature. At liquid helium temperature
the phase coherence relaxation time \emph{$T_2$} reaches $\sim 10$
$\mu$s and, with a Rabi frequency of 20 MHz, yields a single qubit
figure of merit \emph{$Q_M$} of about 500. This shows that a diluted
ensemble of Cr$^{5+}$ (\emph{S} = 1/2) doped K$_3$NbO$_8$ is a
potential candidate for solid-state quantum information processing.
\end{abstract}




\maketitle

Recently, electron spins in solids have been intensively discussed
in terms of physical implementations of a quantum computer
\cite{divincenzo95} and several proposals for embodying solid-state
spin qubits have been put forward. The discussed physical systems
comprise quantum dots \cite{loss98, petta05, koppens06}, phosphorous
donors in silicon \cite{kane98}, endohedral fullerenes
\cite{morton06}, nitrogen-vacancy centers in diamond
\cite{kennedy03, popa04, hanson06, gaebel06, childress06}, molecular
magnets \cite{hill03, ardavan07} and rare-earth ions
\cite{bertaina07}. They commonly make use of the well-characterized
discrete energy levels arising from the spin, orbital, or charge
states. Even though transition metal ions have these states, to our
knowledge, they have not been exploited as a basic building block of
solid-state spin qubits. More importantly, they could be made
essentially free of magnetic anisotropy and therefore suitable for
on-chip deposition and spin manipulation (\emph{vide infra}).

In this Letter, we explore this possibility by lightly doping a
\emph{S} = 1/2 Cr$^{5+}$ ion into the nonmagnetic matrix of
K$_3$NbO$_8$. Cr was chosen since its dominant isotope $^{52}$Cr
(90.5\% natural abundance) has nuclear spin \emph{I} = 0, thus
obviating complications like spin decoherence due to hyperfine
interactions. A further advantage of this system is that, in
principle, it can be isotopically enriched with $^{53}$Cr (\emph{I}
= 3/2, 9.5\% natural abundance) to produce a potential multiqubit
system. We observed Rabi oscillations of the \emph{S} = 1/2 and
\emph{I} = 0 Cr spin in a wide temperature range of 4 to 290 K and
almost independent of field orientation. The measured spin dephasing
times and ease of material synthesis suggest that such spin qubits
based on transition metal oxides might be suitable for scalable
quantum computing.

\begin{figure}
\label{fig:1}
\includegraphics[width=3.2in]{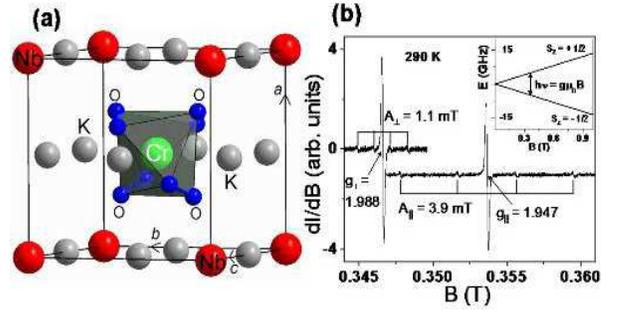}
\caption{(Color online) (a) Crystal structure of Cr$^{5+}$ doped
K$_3$NbO$_8$: Cr [green (or light gray) ball)], Nb [red (or dark
gray) balls], K (gray balls) and O [blue (or black) balls)]. For
clarity, the oxygen atoms are shown only around the Cr ion (dark
gray polyhedron). (b) Room temperature cw EPR spectra (vertically
shifted for clarity) measured at $\nu \sim 9.64$ GHz for
\emph{B}$_0\perp$ \emph{c} and \emph{B}$_0\parallel$\emph{c},
respectively. In both cases, the strong central peak corresponds to
the \emph{S} = 1/2, \emph{I} = 0 resonance and the four weak
sidebands are due to the $^{53}$Cr isotope (\emph{I} = 3/2, 9.5\%
natural abundance). The inset displays the energy level diagram
showing the eigenstates of the \emph{S} = 1/2, \emph{I} = 0 spin
system in an external field.}
\end{figure}

Cr doped K$_3$NbO$_8$ (hereafter Cr:K$_3$NbO$_8$) crystallizes in a
tetragonal unit cell ($I\overline{4}2m$) with lattice parameters
\emph{a} = 6.694 {\AA} and \emph{c} = 7.574 {\AA}. Figure 1(a) shows
a schematic of the Cr:K$_3$NbO$_8$ system. Nb$^{5+}$ ions are shown
at the corners while Cr$^{5+}$ ion is shown at the body center
position. Single crystals of Cr:K$_3$NbO$_8$ were prepared as
described elsewhere \cite{dalal81, cage99} and the Cr$^{5+}$
concentration was determined to be $\sim 0.03\%$ in the studied
sample. Electron paramagnetic resonance (EPR) measurements were
performed using a Bruker Elexsys 680 spectrometer at \emph{X}-band
($\nu \sim 9.64$ GHz) in both continuous-wave (cw) and pulsed modes.
Temperature was varied between 4 and 290 K using helium-flow
cryostats.

The electronic spin Hamiltonian for Cr:K$_3$NbO$_8$ can be written
as follows
\begin{equation}
\label{eq1}
     \hat{H} = \mu_B B_{0} \cdot \textbf{g}\cdot \hat{S} + \hat{H}_{hf} + \hat{H}_{dipole}
\end{equation}
The first term is the electron Zeeman interaction where $\mu_{B}$ is
the Bohr magneton, \textbf{g} the \emph{g}-tensor, and $\hat{S}$ the
spin operator. $\hat{H}_{hf}$ comprises the hyperfine interactions
given by $\sum_{n} \hat{S} \cdot \textbf{A} \cdot \hat{I}^{n}$. Even
though there are no hyperfine interactions with the $^{52}$Cr
(\emph{I} = 0) nuclei, we have found that the superhyperfine
interactions from $^{39}$K nuclei (\emph{I} = 3/2, 93.3\% natural
abundance) are non-negligible and also the interactions with the
$^{93}$Nb (\emph{I} = 9/2, 100\% natural abundance) cannot be ruled
out. The third term represents the dipolar interactions between
electron spins. In solids, dipolar coupling is usually the principle
mechanism limiting the spin-spin relaxation time \emph{$T_2$}.
However, in our case ($\sim 0.03 \%$ Cr$^{5+}$ concentration), a
mean separation of $\sim 8$ nm between Cr$^{5+}$ ions yields an
average electron-electron dipolar interaction of about 0.1 MHz. Our
system is sufficiently diluted that it can be considered as an
almost perfectly isolated Kramer's spin system.

cw EPR spectra of the Cr:K$_3$NbO$_8$ single crystal recorded at
room temperature are presented in Fig. 1(b).  The single central
peak corresponds to the electron-spin transition $\Delta
\emph{S}_{z} = \pm 1$ within the \emph{S} = 1/2, \emph{I} = 0
doublet, as shown in the inset. The \emph{g} values are calculated
as \emph{g}$_{\perp} = 1.9878 \pm 0.0002$ for \emph{B}$_0 \perp$
\emph{c} and \emph{g}$_{\parallel} = 1.9472 \pm 0.0002$ for
\emph{B}$_0\parallel$ \emph{c} and are characteristic of a
tetragonally distorted tetrahedral system with a
3$\emph{d}_{x^2-y^2}$ ground state \cite{cage99}. We note that the
\emph{g} anisotropy in our system is small and therefore can be
treated as quasi-isotropic. The four weak satellite peaks flanking
the central line for both orientations arise from the hyperfine
coupling to the nuclear spin \emph{I} = 3/2 of the $^{53}$Cr isotope
with \emph{A}$_{\parallel} = 3.9 \pm 0.1$ mT and \emph{A}$_{\perp} =
1.1 \pm 0.1$ mT. In this study, we will address only the central
resonance from the \emph{S} = 1/2, \emph{I} = 0 $^{52}$Cr single
qubit spin system.

\begin{figure}
\label{fig:2}
\includegraphics[width=3.2in]{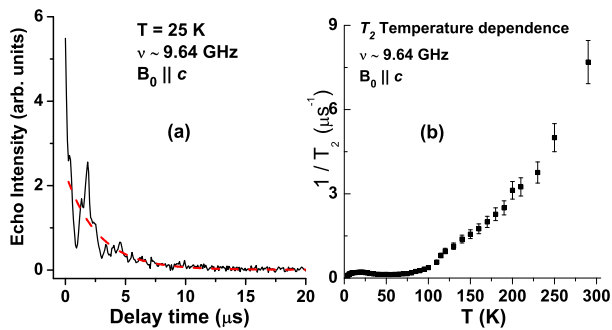}
\caption{(Color online) (a) Hahn echo intensity decay of
Cr:K$_3$NbO$_8$ as a function of delay time at $\nu \sim 9.64$ GHz
and 25 K for \emph{B}$_0 \parallel$ \emph{c}. Observed modulation of
the echo height is due to the superhyperfine coupling with the
$^{39}$K nuclei. Dashed line is the exponential fit to extract
spin-spin relaxation time \emph{$T_2$}. (b) Temperature dependence
of \emph{$T_2$} for \emph{B}$_0 \parallel$ \emph{c} at $\nu \sim
9.64$ GHz.}
\end{figure}

As a first step, the two important characteristics of a qubit
namely, the spin-lattice relaxation time (\emph{$T_1$}) and the
spin-spin relaxation (\emph{$T_2$}) time were measured using the
standard pulse sequences. \emph{$T_1$} was measured by an
inversion-recovery [using free induction decay (FID)] method
employing the sequence $\pi$-$\tau$-$\pi/2$-FID with varying $\tau$.
\emph{$T_1$} increases continually from $\sim 526$ ns at room
temperature to $\sim 1$ s at 4~K. This indicates that the
spin-lattice relaxation is caused by thermal processes. \emph{$T_2$}
is obtained using a 2-pulse Hahn echo decay sequence
$\pi/2$-$\tau$-$\pi$-echo that gives the echo intensity as a
function of $\tau$ [see Fig. 2(a)]. The observed oscillations are
from the electron spin echo envelope modulation (ESEEM) effect.
Their Fourier transform yields the $^{39}$K nuclear spin levels
splittings due to the combined effect of nuclear Zeeman,
superhyperfine and quadrupole interactions. An electron nuclear
double resonance study at 240~GHz \cite{sarita07} yields the
hyperfine couplings which range from 0.41-0.73 MHz and which are
consistent with the ESEEM results. These values allow us to assign
the gaussian lineshape and the linewidth (4.2~MHz) of the EPR
resonance to the unresolved superhyperfine coupling with the
surrounding K nuclei. For a gaussian line, the relation between the
FID decay time \emph {$T^*_2$} and the linewidth $\Delta
\emph{B}_{pp}$ is given as \emph{$T^*_2$} = 2$\sqrt{2}$/$\Delta
\emph{B}_{pp}$. Using the above mentioned $\Delta \emph{B}_{pp}$ of
4.2 MHz, \emph{$T^*_2$} can be estimated as $\sim 108$ ns.

The temperature dependence of \emph{$T_2$} for \emph{B}$_0$
$\parallel$ \emph{c} at $\sim 9.6$ GHz is displayed in Fig. 2(b).
\emph{$T_2$} slowly increases from $\sim 130$ ns at room temperature
to $\sim 10$ $\mu$s at 70 K and essentially remains constant down to
4 K. The details will be discussed in a separate paper
\cite{sarita07}.

\begin{figure}
\label{fig:3}
\includegraphics[width=3.3in]{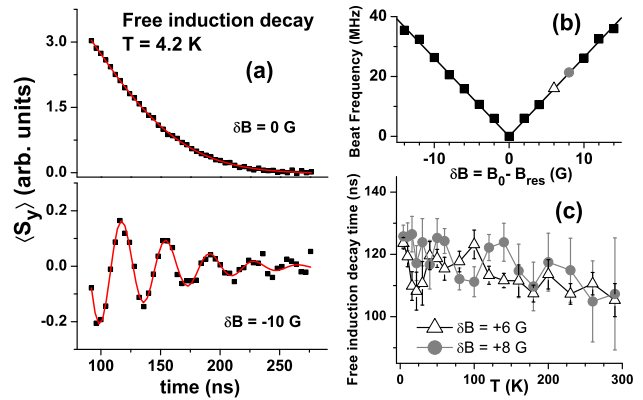}
\caption{(Color online) (a) Cr$^{5+}$ free induction decay (FID) at
4.2 K for two values of applied magnetic field \emph{B}$_0$ =
\emph{B}$_{res}$ + $\delta$\emph{B}, with \emph{B}$_{res}$
determined from the cw resonance experiment for \emph{B}$_0$
$\parallel$ \emph{c} [see Fig. 1(b)]. Experimental data (squares)
are well fitted by a damped oscillatory motion (solid lines). (b)
The beat frequency of the FID versus detuning field. The data are
well described by a through-origin linear fit. (c) The total
dephasing time (\emph{T$_2$*}) as a function of temperature for two
values of $\delta$\emph{B} [shown in (b) by a triangle and a
full circle]. \emph{$T^*_2$} remains practically constant from 4.2 K
to room temperature.}
\end{figure}

The decay of the macroscopic magnetic moment perpendicular to the
magnetic field due to the dephasing of the individual spin packets
with respect to each other is measured by the FID and is shown in
Fig. 3(a). Here, we plot the FID measured at 4.2 K for two detuning
fields $\delta$\emph{B} (= \emph{B}$_0$ - \emph{B}$_{res}$) for
\emph{B}$_0$ $\parallel$ \emph{c}-axis orientation. In a rotating
frame of reference \cite{schweiger01}, \emph{B}$_0$ is along
\emph{z} axis whereas \emph{B}$_1$ is along the \emph{x} axis. A
$\pi/2$ pulse applied with a +\emph{x} phase rotates the spins,
initially aligned along the \emph{z} axis, to the -\emph{y} axis and
when on-resonance (\emph{B}$_0$ = \emph{B}$_{res}$), the spins
remain oriented along the -\emph{y} axis during the free evolution
period and decay with \emph{$T^*_2$}. When the resonance field is
detuned by $\delta$\emph{B}, the average macroscopic magnetization
undergoes a circular motion in the transverse plane with a beat
frequency $\delta$\emph{f}. The decay time and $\delta$\emph{f} were
extracted by fitting the FID data (squares) to a gaussian damped
sinusoidal curve (solid lines). Figure 3(b) shows a plot of
$\delta$\emph{f} versus $\delta$\emph{B}. The through-origin linear
dependence corroborates the relation $\delta$\emph{f} $\propto
\delta$\emph{B}. Figure 3(c) displays the temperature dependence of
\emph{$T^*_2$} for $\delta$\emph{B} = 6 and 8 G. The obtained \emph
{$T^*_2$} of $\sim 115$ ns agrees with the value calculated from the
observed linewidth and remains practically constant from room
temperature to 4.2 K.

\begin{figure}
\label{fig:4}
\includegraphics[width=3.3in]{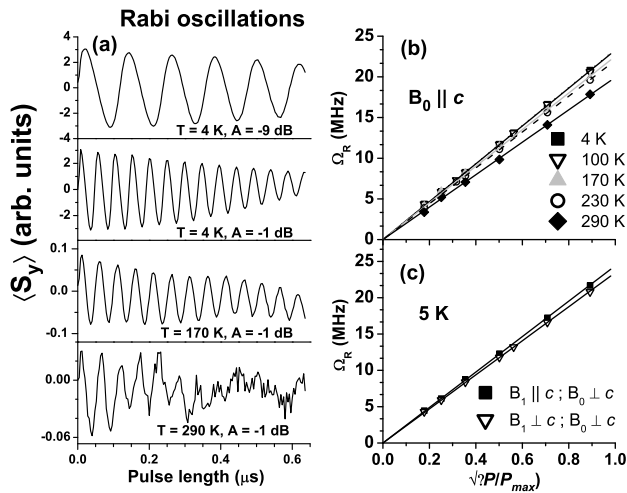}
\caption{(a) Rabi oscillations are observed as time evolution of the
$\langle \emph{S}_y \rangle$ component as a function of pulse
duration at different power levels and temperatures. (b) Linear
dependence of the Rabi frequency ($\Omega_R$) on the microwave
amplitude at 4, 100, 170, 230, and 290 K. (c) Orientation dependence
of $\Omega_R$ measured at 5 K for two microwave field (\emph{B}$_1$)
directions.}
\end{figure}

We have performed driven coherent spin manipulation and obtained the
so-called Rabi oscillations. All measurements were performed by
applying a nutation pulse around +\emph{x} axis of length $\tau$ and
measuring the resulting $\langle \emph{S}_y \rangle$ by the FID
amplitude after a deadtime of $\sim 80$ ns. Several examples of the
observed Rabi oscillations are given in Fig. 4(a) at different power
levels and temperatures. Below 170 K the Rabi amplitude follows the
expected 1/\emph{T} dependence. Above 170 K the additional decrease
of the amplitude (up to 40$\%$ at 290~K) is observed due to the fact
that considerable \emph{$T_2$} decay occurs during the measurement
deadtime. Noticeably, the Rabi oscillations are observable even at
room temperature.

The experimental data are well described by a single exponential
oscillating function
\begin{equation}
\label{eq2}
      \langle S_y \rangle = S_{y(t=0)} e^{-t/\tau_{R}} \sin(\Omega_R
        t).
\end{equation}
The Rabi frequency, $\Omega_R$ and a damping constant, $\tau_R$, are
determined by fitting the data to Eq.(2). In Fig. 4(b) the Rabi
frequency is plotted against the microwave amplitude for
temperatures of 4, 100, 170, 230, and 290 K. Since microwave
amplitude is proportional to the square-root of the incident power,
the \emph{x} axis is taken as $\sqrt{P/P_{max}}$ = 10$^{A/20}$,
where \emph{P} is the power, \emph{P$_{max}$} is the maximum power
($\sim 1$ kW) and \emph{A} is the attenuation in dB. The linear
dependence of $\Omega_R$ on the microwave amplitude is consistent
with what is expected for Rabi oscillations. Since the \emph{g}
value of Cr:K$_3$NbO$_8$ is nearly temperature independent
\cite{cage99}, the smaller slope of the $\Omega_R$ versus incident
power at higher temperatures is due to the smaller \emph{B}$_1$
field at the sample caused by the smaller resonator Q-factor at
higher temperatures. The dependence of $\Omega_R$ on the direction
of the microwave field, \emph{B}$_1$, at 5 K is displayed in Fig.
4(c). We note that ratio of $\Omega_R$ between \emph{B}$_1$
$\parallel$ \emph{c} and \emph{B}$_1$ $\perp$ \emph{c},
($\Omega_{R,\parallel}$) /($\Omega_{R,\perp}$), deviates from the
expected \emph{g}$_\parallel$/\emph{g}$_\perp$ ratio and is likely
due to the small changes in the resonator \emph{Q} factor with the
sample orientation. However, this small directional dependence of
$\Omega_R$ on the microwave field confirms the quasi-isotropic
nature of our system. In addition, we have also checked the
directional dependence of $\Omega_R$ on the external field,
\emph{B}$_0$ and find no difference between the two orientations
(not shown here).

\begin{figure}
\label{fig:5}
\includegraphics[width=3.5in]{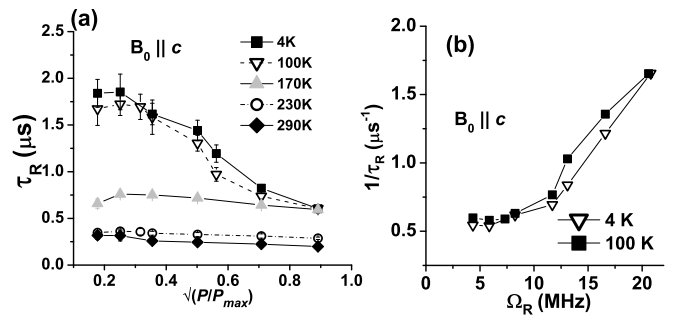}
\caption{(a) Rabi oscillations damping time $\tau_R$ as a function
of microwave amplitude at five temperatures for \emph{B}$_0$
$\parallel$ \emph{c} orientation. (b) Rabi oscillations damping rate
as a function of $\Omega_R$ at 4 K and 100 K for \emph{B}$_0$
$\parallel$ \emph{c} orientation.}
\end{figure}

The variation of Rabi decay time ($\tau_R$) with drive amplitude at
different temperatures is shown in Fig. 5(a). With decreasing
microwave power $\tau_R$ increases and then saturates in the low
power limit regardless of temperature. In addition, the saturation
value of $\tau_R$ increases strongly upon cooling from 290~K to
100~K and then remains more or less constant between 100~K and 4~K.
This behavior resembles the temperature dependence of \emph{$T_2$}
[see Fig. 2(b)], indicating that $\tau_R$ is restricted by a
\emph{$T_2$} mechanism. As in ref. \cite{bertaina07} this can be
explained by a model based on a stochastic field with a normalized
distribution $\beta$ \cite {shakhmuratov97, agnello99},
\begin{equation}
\label{eq3}
       1/\tau_R= \beta\Omega_R + 1/2T_2.
\end{equation}
In our case, this stochastic field can arise from the modulation of
the superhyperfine couplings and the radiation induced changes in
the dipolar field \cite{agnello99}. We find that at high
temperatures $\tau_R \sim$ 2\emph{$T_2$} as expected. In the low
temperature range a marked dependence of the Rabi oscillations decay
rate on the Rabi frequency or the amplitude of the \emph{B}$_1$
field is observed. However, this dependence [see Fig. 5(b)] does not
show the simple linear dependence from Eq.(3) which was observed in
AlO$_4$$^-$ and $E^\prime$ centers in silicate \cite{boscaino93,
agnello99}, and extrapolates to about 2 $\mu$s at temperatures below
100 K, which is somewhat shorter than the observed \emph{$T_2$}
($2.7 - 10$ $\mu$s) in this range [see Fig. 2(b)]. We tentatively
ascribe this to the larger role played by the $^{39}$K
superhyperfine interactions in our system.

The single qubit figure of merit \emph{Q$_M$}, defined as
\emph{Q$_M$} = $\Omega_R$\emph{T$_2$}/$\pi$ \cite{david00},
estimates the efficiency of a quantum device. Using a \emph{$T_2$}
time of 10 $\mu$s we obtain $\sim 500$ coherent single-qubit
operations at liquid helium temperature, implying that Cr$^{5+}$
doped K$_3$NbO$_8$ is a viable electron spin qubit. One way to
improve the \emph{$Q_M$} is to increase the \emph{B}$_1$ by going to
higher microwave power. Even though the \emph{$T_2$} of
Cr:K$_3$NbO$_8$ is smaller than that of nitrogen-vacancy centers in
diamond \cite{hanson06, gaebel06}, it is comparable to that of the
rare-earth qubits \cite{bertaina07} and molecular magnets
\cite{ardavan07}. As mentioned earlier, the spin decoherence in our
system is likely due to the superhyperfine interactions with
$^{39}$K nuclei. This hints a possibility of synthesizing samples
with even longer coherence time. For example, replacement of
$^{39}$K by $^{41}$K reduces the superhyperfine coupling by nearly a
factor of 2. We have also succeeded in synthesizing KCaCrO$_8$,
which should reduce the superhyperfine interactions due to $^{39}$K
by two thirds.

Qubits could be coupled either by dipolar (as in \cite{hanson06}) or
by exchange interactions that are of the order of few Kelvin in this
system \cite{cage01}. We have observed that at concentrations higher
than 5$\%$, we do form Cr-Cr exchange-coupled triplet pairs, which
should in principle enable clustering of several qubits. Such
clusters could be further coupled either by dipolar or exchange
interactions.

For the present studies, sub-mm sized single crystals were
synthesized from solution. In future studies, this technique will be
modified to allow precipitation of nanocrystals in areas designed by
lithography on an electronic circuit. The qubits could be entangled
via photons when integrated in microcavities or via current
oscillations if spin detection is done by electronic transport. It
is important to note that, in contrast to the 2DEG-based quantum
dots where the free electron overlaps over the dot's nuclei, the Cr
spins interact only with the neighboring non-zero spin nuclei of the
crystal. Spin detection can be done in several ways, for instance by
optical detection \cite{berezovsky06}, single-electron transistors
\cite{heersche06}, nanoSQUIDs \cite{cleuziou06}, or microwave
detection by means of SQUID \cite{dalal05} or Hall probes
\cite{loubens06}. The quasi-isotropic magnetic character of the Cr
spins demonstrated here is relevant for most practical
implementations when a controlled positioning of the spins is hard
to achieve. Therefore, we consider this novel material to be
potentially well suited for integration with standard
nanofabrication methods used for on-chip studies.

In conclusion, we have reported on the observation of Rabi
oscillations over $4 - 290$ K in Cr:K$_3$NbO$_8$, an essentially
pure \emph{S} = 1/2, metal-oxide system. We find the intrinsic
phase-coherence time \emph{$T_2$} $\approx 10$ $\mu$s and the single
qubit figure of merit \emph {$Q_M$} $\approx 500$ at liquid helium
temperature. Our results demonstrate that the transition metal
oxide-based spin systems hold high potential for quantum information
applications.

The authors acknowledge the State of Florida, NSF Cooperative
Agreement Grant No. DMR-0084173 and the NSF Grants No. DMR-0520481,
No. NIRT-0506946, NSF-CAREER No. DMR-0645408, No. NHMFL-IHRP-5059,
DARPA-HR0011-07-1-0031 and the Alfred P. Sloan Foundation for
financial support. K.Y.C thanks H. Nojiri for helpful discussions.

\end{document}